\documentclass [11pt]{article}

\ifx\Red\relax\def\Red#1{#1}\fi
\def\Red#1{#1}

\def\textfraction{0.1}

\usepackage{epsfig}
\usepackage{amssymb}
\usepackage{verbatim}
\usepackage{delarray}
\usepackage{graphics}
\usepackage{epic}

\newcommand{\inner}[2]{\langle{#1},{#2}\rangle}

\newcommand{\A}{{\mathcal{A}}}
\newcommand{\B}{{\mathcal{B}}}
\newcommand{\CC}{{\mathcal{C}}}

\newcommand{\I}{{\mathcal{I}}}
\newcommand{\J}{{\mathcal{J}}}

\newcommand{\SSS}{{\mathcal{S}}}
\newcommand{\T}{{\mathcal{T}}}

\newcommand{\C}{{\mathbb{C}}}
\newcommand{\F}{{\mathbb{F}}}
\newcommand{\R}{{\mathbb{R}}}
\newcommand{\Z}{{\mathbb{Z}}}
\newcommand{\zerob}{{\mathbf 0}}
\newcommand{\oneb}{{\mathbf 1}}
\newcommand{\ab}{{\mathbf a}}
\newcommand{\bb}{{\mathbf b}}

\newcommand{\gb}{{\mathbf g}}
\newcommand{\hb}{{\mathbf h}}

\newcommand{\ub}{{\mathbf u}}

\newcommand{\yb}{{\mathbf y}}

\newcommand{\sigmab}{\mbox{\boldmath$\sigma$}}
\newcommand{\Bf}{{\mathfrak{B}}}

\newcommand{\ie}{{\em i.e., }}
\newcommand{\eg}{{\em e.g., }}

\newcommand{\del}{\mathrm{del~}}

\newcommand{\mod}{\mathrm{mod~}}

\newcommand{\openbox}{\leavevmode
     \hbox to.77778em{%
     \hfil\vrule
     \vbox to.675em{\hrule width.6em\vfil\hrule}%
     \vrule\hfil}}
\newcommand{\proofname}{Proof}

\newcommand{\qed}{\hspace*{1cm}\hspace*{\fill}\openbox}

\newtheorem{theorem}{Theorem}

\newtheorem{lemma}[theorem]{Lemma}

\title{Minimal Realizations of Linear Systems:  \\ The ``Shortest Basis" Approach}
\author{G. David Forney, Jr.\footnote{Laboratory for
Information and  Decision Systems,
Massachusetts Institute of Technology,
Cambridge, MA 02139. E-mail: \texttt{forneyd@comcast.net}.}}

\begin{document}
\renewcommand{\textfraction}{0}

\date{}
\maketitle
\thispagestyle{empty}
\begin{center}
\emph{Dedicated to the memory of Ralf Koetter (1963-2009)}
\end{center}
\begin{abstract}
Given a discrete-time linear system $\CC$, a shortest basis for $\CC$ is a set of linearly independent generators for $\CC$ with the least possible lengths.  A basis $\B$ is a shortest basis if and only if it has the predictable span property (\ie has the predictable delay and degree properties, and is non-catastrophic), or alternatively if and only if it has the subsystem basis property (for any interval $\J$, the generators in $\B$ whose span is in $\J$ is a basis for the subsystem $\CC_\J$).  The dimensions of the minimal state spaces and minimal transition spaces of $\CC$ are simply the numbers of generators in a shortest basis $\B$ that are active at any given state or symbol time, respectively.  A minimal linear realization for $\CC$ in controller canonical form follows directly from a shortest basis for $\CC$, and a minimal linear realization for $\CC$ in observer canonical form follows directly from a shortest basis for the orthogonal system $\CC^\perp$.  This approach seems conceptually simpler than that of classical minimal realization theory.
\end{abstract}


\begin{flushright}
It can scarcely be denied that the supreme goal of all theory is to make the \\ irreducible basic elements as simple and as few as possible.---  A. Einstein \cite{Einstein}\footnote{In other words, everything should be made as simple as possible, but not simpler.}
\end{flushright}

\normalsize


\section{Introduction}

The minimal realization problem of linear system theory is the problem of finding a state-space realization for a given linear system, often time-invariant, that has the smallest possible state space(s), possibly in some predetermined canonical form.  The problem becomes nontrivial in the general case of multivariable and/or time-varying linear systems.  The system is usually specified by its impulse response(s), or by some realization that may be nonminimal.  

This problem has been studied since the rise of the state-space paradigm in the early 1960s.  The classical solution to this problem is usually expressed by the mantra ``minimal = controllable + observable."  Many concrete algorithms have been developed to solve it, typically making heavy use of linear algebra and matrix manipulations;  see \eg \cite{D00}. 

There is a much simpler approach to the minimal realization problem, at least conceptually;  namely, the ``shortest basis" approach, as we shall call it here.  For linear time-invariant systems, this approach was developed in \cite{F75}.  It has been used extensively in the literature of minimal trellis (state-space) realizations of linear block codes \cite{KS95, V98}, where shortest bases are called ``trellis-oriented" \cite{F88} or ``minimum-span" generator matrices.  Analogous results are developed in a very general group-theoretic setting in \cite{FT93}.  This paper may be regarded either as a specialization of  \cite{FT93} to the case of linear systems over fields, or, preferably (because all proofs are linear-algebraic), a generalization of \cite{F75} to time-varying linear systems, or of \cite{KS95} to infinite-time-axis linear systems.

Much of this paper is a tutorial overview of known results, aimed particularly at a system-theory audience.  The following two results are new, as far as we know:
\begin{itemize}
\item $\B$ is a shortest basis if and only if it has the predictable span property;
\item $\B$ is a shortest basis if and only if it has the subsystem basis property.
\end{itemize}
But many of these results are not very well known, at least in system theory, and many of the proofs are new.



\section{Preliminaries}

We  focus on \emph{linear discrete-time systems} over a field $\F$.  In system theory, $\F$ is usually the real field $\R$ or the complex field $\C$, whereas in coding theory $\F$ is usually a finite field.  The astute reader will notice that most proofs depend only on the group property of linear systems, and therefore apply more generally to group systems.  The reader will also observe that our approach is entirely algebraic, and that analytical issues such as stability and convergence play no role.

A discrete-time system has a discrete, ordered \emph{time axis} $\I$, which we take to be the set of integers $\Z$, or a subinterval of $\Z$.  We  use notation such as $[n, m) = \{k \in \Z : n \le k < m\}$ for subintervals of $\Z$.

Here, as in behavioral system theory \cite{W86}, a system will be defined by the set $\CC$ of all of its possible \emph{trajectories} $\ab = \{a_k : k \in \I\}$ (its ``behavior"), where each \emph{symbol} $a_k$ lies in some \emph{alphabet} $A_k$.  If the system is \emph{linear} over a field $\F$, then each alphabet $A_k$ is a vector space over $\F$, assumed to be finite-dimensional, and $\CC$ is a subspace of the Cartesian-product vector space $\A = \prod_{k \in \I} A_k$. 

If all alphabets $A_k$ are equal to $\F$, then a trajectory $\ab$ may be represented by its $z$-transform $a(z) = \sum_k a_k z^{-k}$, as in linear system theory, or by its $D$-transform $a(D) = \sum_k a_k D^k$, as in coding theory.  The subtle difference (apart from the obvious difference $D = z^{-1}$) is that $D$ is simply an indeterminate, whereas $z$ is often regarded as a complex variable.  We shall use both kinds of transforms in this paper, but we shall always regard $z$ as simply an indeterminate.

The \emph{support} of a trajectory $\ab$ is the subset of all indices $k \in \I$ such that $a_k \neq 0$.  A trajectory is \emph{zero, finite} or \emph{infinite} according to whether the size of its support is zero, finite or infinite. 

 If the support of a nonzero trajectory $\ab$ has a minimum element $k_{\min}$, then $\ab$ is called \emph{Laurent}, and $k_{\min}$ is called the \emph{delay} of $\ab$, denoted by $\del \ab = k_{\min}$.  By convention, the zero trajectory is defined as Laurent, and its delay is defined as $\del \zerob = +\infty$.  
 
 In the body of this paper, we shall require all trajectories to be Laurent, as is common in coding theory.
This restriction simplifies our exposition and yields a symmetrical duality theory, but forecloses consideration of uncontrollable systems with autonomous components.  In an appendix, we show that the extension of our approach to uncontrollable/autonomous systems is straightforward. 

 If the support of $\ab$ has a maximum element $k_{\max}$, then $k_{\max}$ is called the \emph{degree} of $\ab$, denoted by $\deg \ab = k_{\max}$.  By convention, the degree of the zero trajectory is defined as $\deg \zerob = -\infty$. 

The set of all $z$-transforms $a(z)$ or $D$-transforms $a(D)$ of Laurent trajectories $\ab$ over $\F$ on the time axis $\I = \Z$ is called the set of all \emph{formal Laurent series} in $\F$ over $z^{-1}$ or $D$, and is conventionally denoted by $\F((z^{-1}))$ or $\F((D))$, respectively.  A nice algebraic property of $\F((z^{-1}))$ or $\F((D))$ is that it forms a field, with multiplication defined by sequence convolution.  In particular, every nonzero $a(z) \in \F((z^{-1}))$ has a Laurent inverse $1/a(z)$, which may be computed by long division.

The set of all $z$-transforms $a(z)$ or $D$-transforms $a(D)$ of finite trajectories $\ab$ over $\F$ with $\del \ab \ge 0$ is called the set of all \emph{polynomials} in $\F$ over $z^{-1}$ or $D$, and is denoted by $\F[z^{-1}]$ or $\F[D]$, respectively.  For polynomials, our definition of ``degree" coincides with the standard definition.

A \emph{linear time-invariant} (LTI) system is a linear system $\CC$ whose time axis is $\I = \Z$, whose alphabets $A_k$ are all equal, and which satisfies $D\CC = \CC$, where $D$ is the \emph{delay operator} that transforms $\ab \in \CC$ to $D\ab = \{a_{k-1}: k \in \I\}$. 
 (This usage of $D$ is compatible with that in $D$-transforms, since if the $D$-transform $\ab$ is $a(D)$, then that of $D\ab$ is $Da(D)$.)  This implies that if $\ab \in \CC$, then every positive or negative shift $D^k\ab, k \in \Z$, is in $\CC$.  Note that the set $\F((D))$ of all formal Laurent series over $\F$ is time-invariant.

It is natural to define a linear system $\CC$ by a linearly independent set $\gb$ of \emph{generators}, called a \emph{basis}, such that every trajectory in $\CC$ is a unique linear (over $\F$) combination of the generators.  We say that a generator is \emph{involved in} a linear combination if it has a nonzero coefficient in that combination.      For an LTI system, it is natural to choose a basis that consists of all the shifts $D^k\gb_j, k \in \Z$, of a set  $\{\gb_j\}$ of \emph{fundamental generators} $\gb_j$.


\vspace{1ex}
\noindent
\textbf{Example 1} (single-input, single-output LTI system).  Consider a real or complex discrete-time linear filter whose impulse response has $z$-transform $g(z) = 1/(1 - \beta z^{-1})$, which denotes the Laurent $z$-transform $1 + \beta z^{-1} + \beta^2 z^{-2} + \cdots$.\footnote{We need not restrict $|\beta| < 1$ if we are not concerned with issues of stability or convergence.}  What is the set $\CC$ of trajectories associated with this filter?  We might say that $\CC$ is the set of all output sequences of the filter in response to all Laurent input sequences.  But then $\CC$ would simply be the set of all Laurent sequences, since every Laurent sequence $a(z)$ could be the output sequence if the input were the Laurent sequence $a(z)(1 - \beta z^{-1})$;  so such a definition would fail to capture the particular characteristics of this filter.  Therefore we instead define the set $\CC$ of trajectories of this system as the set of all input-output  pairs as the input runs through all Laurent sequences: 
$$
\CC = \{(u(z), u(z)g(z)) : u(z) \in \F((z^{-1}))\}.
$$
$\CC$ is evidently an LTI system.
The set of all shifts of the fundamental input-output pair $(1, g(z))$ is a basis for $\CC$. \qed

\vspace{1ex}
\noindent
\textbf{Example 2} (binary linear block code).  The $(8,4,4)$ first-order binary Reed-Muller code is the four-dimensional subspace $\CC$ of  $(\F_2)^8$ that comprises all 16 binary 8-tuples that can be obtained as binary linear combinations of the following four generators:
\begin{eqnarray*}
\gb_1 & = & 1111 0000; \\
\gb_2 & = & 1100 1100; \\
\gb_3 & = & 1010 1010; \\
\gb_4 & = & 1111 1111. 
\end{eqnarray*}
$\CC$ may be regarded as a linear system over the binary field $\F_2$ that is defined on the finite time axis $\I = [0,8)$, with the basis given above.  Of course, a system defined on a finite time axis cannot be time-invariant. \qed

\vspace{1ex}

Given a linear system $\CC \subseteq \A = \Pi_{k \in \I} A_k$ defined on a time axis $\I$, a (state-space) \emph{realization} of $\CC$ (sometimes called a ``trellis realization" in coding theory) will be defined in terms of:
\begin{itemize}
\item  A \emph{state time axis} $\I_S \subseteq \Z$, such that symbol time $k \in \I$ occurs \emph{between} state time $k \in \I_S$ and state time $k+1 \in \I_S$.    If $\I = \Z$, we take $\I_S = \Z$;  but if $\I = [0, n)$, we take $\I_S = [0, n]$.
\item  A set of \emph{state spaces} $\Sigma_k, k \in \I_S$. 
\item  For each $k \in \I$, a set $\T_k$ of allowable \emph{transitions} $(\sigma_k, a_k, \sigma_{k+1}) \in \Sigma_k \times A_k \times \Sigma_{k+1}$.
\end{itemize}
The \emph{full behavior} $\Bf$ of the realization is then the set of all symbol-state trajectories $(\ab, \sigmab) \in \A \times \Pi_{k \in \I_S} \Sigma_k$ such that $(\sigma_k, a_k, \sigma_{k+1}) \in \T_k$ for all $k \in \I$.  The system $\CC$ realized by the realization is the set of all symbol trajectories $\ab \in \A$ that appear in some symbol-state trajectory $(\ab, \sigmab) \in \Bf$.

A realization is \emph{linear} if all symbol alphabets $A_k$ and state spaces $\Sigma_k$ are vector spaces over a field $\F$, and the \emph{transition spaces} $\T_k$ (``local behaviors") are subspaces of the vector spaces $\Sigma_k \times A_k \times \Sigma_{k+1}$ for all $k \in \I$.  A realization is \emph{minimal} if the state spaces $\Sigma_k$ are as small as possible for all $k \in \I_S$.  We will see that every linear system $\CC$ has a realization that is both linear and minimal.

\section{Minimal state and transition spaces}

In this section we recapitulate well-known results about minimal state spaces, and less well-known results about minimal transition spaces.

\subsection{Minimal state spaces}

A fundamental result of Willems' behavioral system theory \cite{W86, W89} is that, given a linear system $\CC$, the minimal state space at each possible state time is unambiguously defined.  

A state space $\Sigma_k$ at state time $k \in \I_S$ may be considered to be defined by a \emph{cut} of the symbol time axis $\I$ between symbol time $k-1$ and symbol time $k$.  Such a cut partitions $\I$ into two disjoint subintervals: a \emph{past} $k^- = \{k' \in \I : k' <  k\}$ and a \emph{future} $k^+ = \{k' \in \I : k' \ge k\}$.  

The fundamental property of states is the \emph{Markov property}:  the future should be conditionally independent of the past, given the state.  In a state-space realization, this translates to a requirement that two symbol trajectories up to time $k-1$ may arrive at the same state in $\Sigma_k$ if and only if the sets of their possible future continuations from time $k$ on are identical.\footnote{In automata theory, this is called \emph{Nerode equivalence}.} 

In the linear case, it is easy to identify when this happens.  Let  the \emph{past subsystem} $\CC_{k^-}$ and \emph{future subsystem} $\CC_{k^+}$ be defined as the subsets of $\CC$ that are all-zero on the future $k^+$ and the past $k^-$, respectively.  Both $\CC_{k^-}$ and $\CC_{k^+}$ are evidently linear subsystems of $\CC$.  Then the \emph{minimal state space} at state time $k$ is the following quotient space:
$$
\Sigma_k = \frac{\CC}{\CC_{k^-} \times \CC_{k^+}}.
$$

The proof is essentially as follows.  The quotient space $\CC/(\CC_{k^-} \times \CC_{k^+})$ is a disjoint union of cosets $\ab + (\CC_{k^-} \times \CC_{k^+})$, where each coset representative $\ab$ is a trajectory in $\CC$.  
Define $P_{k^-}:  \A \to \A$ and $P_{k^+}:  \A \to \A$ as the projection operators onto the past $k^-$ and future $k^+$, respectively.  The coset $\ab + (\CC_{k^-} \times \CC_{k^+})$ is then precisely the
Cartesian product \pagebreak
$$
\ab + (\CC_{k^-} \times \CC_{k^+}) = 
\left( P_{k^-}(\ab) + \CC_{k^-} \right) \times \left( P_{k^+}(\ab) + \CC_{k^+} \right);
$$
\ie the set of trajectories in $\CC$ whose past projections are in the coset $P_{k^-}(\ab) + \CC_{k^-}$ of $\CC_{k^-}$, and whose future projections are in the coset $P_{k^+}(\ab) + \CC_{k^+}$ of $\CC_{k^+}$.  Since every such past projection has the same set of future continuations, all of these trajectories may pass through the same state at time $k$.  On the other hand, any past projection in any other coset may not pass through the same state at time $k$, since it has a disjoint set of future continuations.  Thus any minimal state space must be in one-to-one correspondence with this set of cosets;  \ie with $\Sigma_k$.

By a simple extension of the above argument,  or by an elementary result in group theory (the first theorem about subdirect products in $\cite{Hall}$), the minimal state space $\Sigma_k$ is also isomorphic to the following quotient spaces, called the ``past-induced" and ``future-induced" state spaces:
$$
\Sigma_k \simeq \frac{P_{k^-}(\CC)}{\CC_{k^-}} \simeq \frac{P_{k^+}(\CC)}{\CC_{k^+}},
$$
where $P_{k^-}(\CC)$ and $P_{k^+}(\CC)$ are the sets of past and future projections of $\CC$, respectively.

It is straightforward to define a linear state-space realization of a linear system $\CC$ that uses the minimal state spaces $\Sigma_k$ for every time $k$.  Let each trajectory $\ab \in \CC$ pass through the sequence of states $\sigma_k(\ab) \in \Sigma_k$ for all $k$ that are defined by the natural maps from $\CC$ to the quotient spaces $\CC/(\CC_{k^-} \times \CC_{k^+})$.  This is called the \emph{canonical realization} of a linear system $\CC$.  Among other things, the canonical realization shows that there exist linear realizations whose state spaces are minimal at every time $k \in \I_S$.

We shall require that all minimal state spaces $\Sigma_k$ be finite-dimensional.  As we shall see shortly, this condition ensures that only finitely many generators in a shortest basis $\B$ will be ``active" at any state time $k$, which in turn ensures that the number of generators that affect any component $a_k$ of any trajectory $\ab \in \CC$ will be finite, even when $\B$ is infinite.  

\subsection{Minimal transition spaces}

We now discuss minimal transition spaces.  In coding theory, transition spaces (``branch spaces," ``trellis sections," ``local constraint codes") have come to be regarded as having importance equal to or possibly even greater than that of state spaces.


Minimal transition spaces are characterized by the following theorem: 


\begin{theorem}[Minimal transition spaces]
In any minimal realization of a linear system $\CC$, for every symbol time $k \in \I,$ the set of transitions is in one-to-one correspondence with the following quotient space, called the \emph{minimal transition space} at symbol time $k$:
$$
\T_k = \frac{\CC}{\CC_{k^-} \times \CC_{(k+1)^+}}.
$$
\end{theorem}


\noindent
\textit{Proof}.  In a minimal realization, the state spaces $\Sigma_k$ and $\Sigma_{k+1}$ are in one-to-one correspondence to $P_{k^-}/\CC_{k^-}$ and $P_{(k+1)^+}/\CC_{(k+1)^+}$, respectively.  The set of all trajectories $\ab \in \CC$ that pass through a given transition $(\sigma_k, a_k, \sigma_{k+1}) \in \Sigma_k \times A_k \times \Sigma_{k+1}$ is the set that have a past projection $P_{k^-}(\ab)$ in the coset of $\CC_{k^-}$ that corresponds to $\sigma_k$, a time-$k$ projection $P_{\{k\}}(\ab)$ equal to $a_k$, and a future projection $P_{(k+1)^+}(\ab)$ in the coset of $\CC_{(k+1)^+}$ that corresponds to $\sigma_{k+1}$. Thus the trajectories of $\CC$ that pass through the same transition at symbol time $k$ are precisely those that lie in the same coset of $\CC_{k^-} \times \CC_{(k+1)^+}$, so the set of transitions corresponds one-to-one to the set $\T_k$ of cosets of $\CC_{k^-} \times \CC_{(k+1)^+}$ in $\CC$. \qed \pagebreak \vspace{1ex}

It is easy to see that in a canonical realization, each trajectory $\ab \in \CC$ passes through the sequence of transitions that are defined by the natural maps from $\CC$ to the quotient spaces $\CC/(\CC_{k^-} \times \CC_{{k+1}^+})$, where $\sigma_k(\ab)$ and $\sigma_{k+1}(\ab)$ are identified with past-induced and future-induced states, respectively.

\section{Shortest bases and minimal realizations}

It is convenient to have a basis for a linear system $\CC$ from which a minimal state-space realization of $\CC$ and its parameters can be read directly, ``by inspection."  In this section we will see that a shortest basis has these properties.  In the literature of minimal trellis realizations of linear block codes, such a shortest basis is called a ``trellis-oriented"  or ``minimum-span" generator matrix \cite{KS95, V98}.  In system theory terms, it yields a minimal realization of $\CC$ in ``controller canonical form" \cite{Kailath}.

We first mention some technical ``well-behavedness" requirements that we  impose on $\CC$ for simplicity.  Let $\CC_\mathrm{finite}$ denote the set of all finite trajectories in $\CC$.  We say that a linear combination of trajectories is \emph{Laurent} if it involves only trajectories whose delays are not less than some minimum delay $k_{\min}$;  then the combination is a Laurent trajectory with delay $\ge k_{\min}$.  We require that $\CC$ be the set of all Laurent linear combinations of $\CC_\mathrm{finite}$.  Then $\CC$ is generated by its finite trajectories;  such a linear system is called \emph{controllable} \cite{W89, F97}.
Also, $\CC$ is then complete \cite{W89} in a Laurent sense;  indeed, $\CC$ is the Laurent completion of $\CC_\mathrm{finite}$.
  For an extension of our development to uncontrollable systems, see the Appendix, or \cite{FT93}. 
  
\subsection{Shortest bases}

The \emph{span} of a nonzero finite trajectory $\ab \in \A$ is the shortest interval that contains its support, namely $[\del \ab, \deg \ab]$, and the \emph{length} of $\ab$ is the size of its span, namely $\deg \ab - \del \ab + 1$.

Loosely, a \emph{shortest basis} $\B$ for a linear system $\CC$ will be defined as a basis whose elements are as short as possible.  We will see as we proceed that this concept is well defined.  


\vspace{1ex}
\noindent
\textbf{Example 1} (cont.). Recall that the single-input, single-output linear time-invariant system of Example 1 is defined as  the set $\CC$ of all input-output  pairs
$$
\CC = \{u(z) (1, g(z)) : u(z) \in \F((z^{-1}))\}
$$
 as the input $u(z)$ runs through all Laurent sequences,
where $g(z) = 1/(1 + \beta z^{-1})$. 
In other words, $\CC$ is the set of all multiples of the basic input-output pair $(1, g(z))$ by sequences in the Laurent field $\F((z^{-1}))$.  It is easy to see that a nonzero trajectory in $\CC$ is finite if and only if the corresponding input sequence $u(z)$ is a finite multiple of $1/g(z) = 1 + \beta z^{-1}$.  Hence the set $\CC_\mathrm{finite}$ of finite trajectories in $\CC$ is the set of multiples of the finite pair $(1 + \beta z^{-1}, 1)$ by finite Laurent sequences $v(z)$:
$$
\CC_\mathrm{finite} = \{(v(z)(1 + \beta z^{-1}), v(z)) : v(z) \in (\F((z^{-1})))_\mathrm{finite}\},
$$
From this it is clear that the shortest trajectories in $\CC$ have length 2, and that the set of shifts of the length-2 trajectory $(1 + \beta z^{-1}, 1)$ is a shortest basis $\B$ for $\CC$. \qed

\vspace{1ex}
\noindent
\textbf{Example 2} (cont.).  Recall that the $(8,4,4)$ binary Reed-Muller code of Example 2 is the set $\CC$ of all 16 binary linear combinations of the four generators $(11110000, 11001100, 10101010,$ $11111111)$.
By examining all 16 codewords, we find that the shortest nonzero codewords are the three length-4 8-tuples $1111 0000, 0011 1100$ and $0000 1111$, which are evidently independent.  The next-shortest codeword that is independent of the previous three is the length-6 8-tuple $0101 1010$.  Since the dimension of $\CC$ is 4, these four 8-tuples comprise a shortest basis $\B$ for $\CC$. \qed \pagebreak \vspace{1ex}

A shortest basis $\B$ may be found by the following greedy construction.  Recall that if $\J$ is a subinterval of the time axis $\I$, then $\CC_\J$ denotes the \emph{subsystem} of $\CC$ consisting of all trajectories $\ab \in \CC$ whose span is contained in $\J$.  First, for every length-1 subinterval $\J \subseteq \I$, find a set of linearly independent length-1 generators for $\CC_\J$.  Next, for every length-2 subinterval $\J \subseteq \I$, find a minimal set of additional independent length-2 generators sufficient (in combination with the previous length-1 generators) to generate all length-2 trajectories in $\CC_\J$.  And so forth. 

Assuming that $\CC$ is controllable, this algorithm will eventually find a shortest basis $\B$ of finite linearly independent generators for $\CC$.  Furthermore, it is clear that any shortest basis for $\CC$ may be constructed in this way.  Since $\dim~ \CC_\J$ is a parameter of the system $\CC$ for every subinterval $\J \subseteq \I$,  it follows that the set of lengths of generators in any shortest basis $\B$ for $\CC$ is the same. 


\subsection{The predictable span property and the subsystem basis property}

We  now introduce two properties, the predictable span property and the subsystem basis property, and show that a basis $\B$ is a shortest basis if and only if it has either of these properties.

We note that linear independence implies that every trajectory $\ab \in \CC$ has a unique expression as a (possibly infinite) linear combination of generators in $\B$, so that we may speak of the \emph{generators of} $\ab$, meaning the subset $\SSS(\ab)$ of generators of $\B$ that are involved in this unique linear combination.

If  $\SSS(\ab)$ is the set of generators of a trajectory $\ab \in \CC$, then $\del \ab \ge k_{\min} = \min_{\gb \in \SSS(\ab)} \del \gb$, and $\deg \ab \le k_{\max} = \max_{\gb \in \SSS(\ab)} \deg \gb$, where strict inequality may occur due to cancellations.  A basis $\B$ has the \textbf{predictable span property} (PSP) if inequality never occurs;  \ie if the span of $\ab$ is always equal to $[k_{\min}, k_{\max}]$.

For finite linear combinations, $\B$ evidently has the PSP if and only if it has the \emph{predictable delay property} (\ie  $\del \ab = k_{\min}$ always) and the \emph{predictable degree property} (\ie  $\deg \ab = k_{\max}$ always) \cite{F75}.  Clearly $\B$ has the predictable delay property if and only if the time-$k$ symbols of the delay-$k$ generators in $\B$ are linearly independent, so cancellation can never occur;  similarly $\B$ has the predictable degree property if and only if the time-$k$ symbols of the degree-$k$ generators in $\B$ are linearly independent.

These linear independence properties have an immediate corollary:

\begin{lemma}[A finite number of generators start and stop at each time]
If a basis $\B$ for $\CC$ has the predictable delay (resp.\ degree) property, then the number of generators in $\B$ that have delay $k$ (resp.\ degree $k$) is not greater than $\dim A_k$. 
In particular, if $\dim A_k = 1$, then at most one generator in $\B$ can start or stop at time $k$.
\end{lemma}
\noindent
\emph{Proof}.  If $\B$ has the predictable delay property, then the set of time-$k$ symbols of delay-$k$ generators in $\B$ is a linearly independent subset of elements of the time-$k$ symbol alphabet $A_k$; similarly the set of time-$k$ symbols of degree-$k$ generators in $\B$ is a linearly independent subset of $A_k$. \qed \vspace{1ex}

Now let us consider infinite Laurent linear combinations.  Recall that a linear combination of generators is \emph{Laurent} if it involves only generators whose delays are not less than some finite minimum delay $k_{\min}$.  On the other hand, we must have $k_{\max} = \infty$, since an infinite number of generators are involved and there can only be a finite number of each finite degree, under our assumption that $\dim A_k$ is finite.  Therefore $\B$ has the PSP for infinite linear combinations if and only if it has the predictable delay property and  all infinite linear combinations are infinite.  (See Section V for examples of infinite linear combinations that are finite.)

Borrowing a term from coding theory, we  say that a basis $\B$ for $\CC$ is \emph{catastrophic} if there exists a finite $\ab \in \CC$ that is equal to an infinite linear combination of generators.  Thus, in summary, $\B$ has the PSP if and only if $\B$ has the predictable delay and degree properties, and $\B$ is non-catastrophic.

Secondly, we will say that a basis $\B$ has the \textbf{subsystem basis property}  (SBP) if 
for any subinterval $\J$ of the time axis $\I$, the set of generators in $\B$ whose span is contained in $\J$ is a basis for the subsystem $\CC_\J$.
By construction, a shortest basis has this property for all finite $\J$.

Now we show that a basis $\B$ is a shortest basis if and only if it has the PSP, or the SBP: 

\begin{theorem}[Shortest basis $\Leftrightarrow$ PSP $\Leftrightarrow$ SBP]
For a basis $\B$ of a controllable linear system $\CC$, the following are equivalent:
\begin{enumerate}
\item $\B$ has the predictable span property;
\item $\B$ has the subsystem basis property;
\item $\B$ is a shortest basis for $\CC$.
\end{enumerate}
\end{theorem}
\noindent
\emph{Proof}.  ($1 \Rightarrow 2$)  On the one hand, a linear combination of generators in $\B$ whose span is contained in $\J$ must be a trajectory $\ab \in \CC_\J$. On the other hand, if $\B$ has the PSP and $\ab \in \CC_\J$, then the minimum degree of the generators of $\ab$ is $\del \ab \in \J$, and the maximum degree is $\deg \ab \in \J$, so every $\ab \in \CC_\J$ is a linear combination of generators in $\B$ whose span is contained in $\J$. \vspace{1ex}

\noindent
($2 \Rightarrow 3$) If $\B$ has the SBP, then,  for each finite subinterval $\J$, the generators in $\B$ whose span is precisely $\J$ could be chosen in the shortest-basis construction process, so $\B$ is a shortest basis.  \vspace{1ex}

\noindent
($3 \Rightarrow 1$) If $\B$ is a shortest basis for $\CC$, then the set of time-$k$ symbols of delay-$k$ generators must be a linearly independent subset of the time-$k$ symbol alphabet $A_k$, else there would be a finite linear combination $\bb$ of delay-$k$ generators with $\del \bb > k$, and with $\deg \bb \le k_{\max} = \deg \gb$, the degree of a greatest-degree generator $\gb$ involved in this linear combination, so $\bb$ would be shorter than $\gb$, and $\bb$ could replace $\gb$ in $\B$ to produce a shorter basis $\B'$;  contradiction.  So $\B$ must have the predictable degree property.  Similarly, $\B$ must have the predictable delay property.  Finally, by the shortest-basis construction, every finite trajectory $\ab \in \CC$ is uniquely expressible as a finite linear combination of finite generators in $\B$, so $\B$ must be non-catastrophic.  \qed \vspace{1ex}

\subsection{Dimensions of minimal state and transition spaces}

We now show how to determine the dimension of the minimal state space $\Sigma_k = \CC/(\CC_{k^-} \times \CC_{k^+})$ ``by inspection" from any shortest basis $\B$ for a linear system $\CC$, for any cut of the time axis $\I$ into a past subset $k^-$ and a future subset $k^+$.  


We partition the generators in $\B$ into three subsets:  a past subset $S_{k^-}$ consisting of those generators in $\B$ whose support is contained in $k^-$, a future subset $S_{k^+}$  consisting of those generators in $\B$ whose support is contained in $k^+$, and a remainder subset $R_k$ consisting of the remaining generators, which we  call the \emph{active generators at state time $k \in \I_S$}.  By the subsystem basis property of shortest bases, $S_{k^-}$ is a basis for $\CC_{k^-}$ and $S_{k^+}$ is a basis for $\CC_{k^+}$.

\begin{theorem}[Minimal state space dimension]
For any state time $k \in \I_S$, the dimension of the minimal state space $\Sigma_k$ of a linear system $\CC$ is the number of active generators at state time $k$ in any shortest basis $\B$ for $\CC$.
\end{theorem}

\noindent
\textit{Proof}.  Since $\B$ is a basis for $\CC$, $S_{k^-}$ is a basis for $\CC_{k^-}$, and $S_{k^+}$ is a basis for $\CC_{k^+}$, the quotient space $\Sigma_k = \CC/(\CC_{k^-} \times \CC_{k^+})$ has a basis consisting of the remaining linearly independent generators in the remainder subset $R_k$, and thus has dimension $|R_k|$. \qed \vspace{1ex}

\noindent
\textbf{Example 1} (cont.). Recall that a shortest basis for the system $\CC$ of Example 1 is  the set of shifts of the length-2 trajectory $(1 + \beta z^{-1}, 1)$. For any state time $k \in \Z$, precisely one of these generators is active;  for example, at state time 1, only the fundamental generator $(1 + \beta z^{-1}, 1)$ is active, since all of its shifts have supports either in the past $1^- = \{k < 1\}$ or in the future $1^+ = \{k \ge 1\}$.  Thus the dimension of the minimal state space $\Sigma_k$ of $\CC$ at any state time $k \in \Z$ is 1.

More generally, let $\CC$ be any single-input, single-output LTI system consisting of all input-output trajectories $\{u(z)(1, g(z)) : u(z) \in \F((z^{-1}))\}$, where $g(z)$ is a causal rational impulse response $g(z) = a(z)/b(z)$ with $a(z), b(z)$ being relatively prime polynomials in $\F[z^{-1}]$, with $\del a(z) \ge 0$ and $\del b(z) = 0$.  Then a shortest basis for $\CC$ is the set of all shifts of the fundamental input-output trajectory $(b(z), a(z))$, whose delay is 0 and whose degree is $\delta = \max\{\deg b(z), \deg a(z)\}$.  Precisely $\delta$ of these shifts are active at any state time $k \in \Z$;  therefore the dimension  of the minimal state space $\Sigma_k$ of $\CC$ at any state time $k \in \Z$ is $\delta$.
\qed

\vspace{1ex}
\noindent
\textbf{Example 2} (cont.).  Recall that the $(8,4,4)$ code of Example 2 has the following shortest basis:
\begin{eqnarray*}
\gb_1 & = & 1111 0000; \\
\gb_2 & = & 0011 1100; \\
\gb_3 & = & 0000 1111; \\
\gb_4 & = & 0101 1010. 
\end{eqnarray*}
Notice that the generators in this set ``start" at symbol times $0, 1, 2$ and 4, and ``stop" at symbol times $3, 5, 6$ and 7.  The numbers of generators that are active at state times $0, 1, \ldots, 8$ are therefore $0, 1, 2, 3, 2, 3, 2, 1, 0$, respectively, which are therefore the minimal state space dimensions at these times.  We say that the \emph{state-space dimension profile} of $\CC$ is $\{0, 1, 2, 3, 2, 3, 2, 1, 0\}$.
\qed \vspace{1ex}

Similarly, from the minimal transition space theorem, we can determine the dimension of the minimal transition space $\T_k$ from any shortest basis $\B$ for a linear system $\CC$.  We now partition the time axis $\I$ into three subintervals: a past interval $k^-$, the time $\{k\}$, and a future interval $(k+1)^+$.  We partition the generators in $\B$ into three subsets:  a past subset $S_{k^-}$ consisting of those generators in $\B$ whose support is contained in $k^-$, a future subset $S_{(k+1)^+}$  consisting of those generators in $\B$ whose support is contained in $(k+1)^+$, and a remainder subset $T_k$ consisting of the remaining generators, which we  call the \emph{active generators at symbol time $k \in \I$}.  


\begin{theorem}[Minimal transition space dimension]
For any symbol time $k \in \I$, the dimension of the minimal transition space $\T_k$ of a linear system $\CC$ is the number of active generators at symbol time $k$ in any shortest basis $\B$ for $\CC$.
\end{theorem}

\noindent
\textit{Proof}.  Since $\B$ is a basis for $\CC$, $S_{k^-}$ is a basis for $\CC_{k^-}$, and $S_{(k+1)^+}$ is a basis for $\CC_{(k+1)^+}$, the quotient space $\T_k = \CC/(\CC_{k^-} \times \CC_{(k+1)^+})$ has a basis consisting of the remaining linearly independent generators in the remainder subset $T_k$, and thus has dimension $|T_k|$. \qed \vspace{1ex}

\noindent
\textbf{Example 1} (cont.). For the system $\CC$ of Example 1, we may take $\B$ as the set of shifts of the length-2 trajectory $(1 + \beta z^{-1}, 1)$. For any symbol time $k \in \Z$, precisely two of these generators are active;  for example, at symbol time 1, the generators $(1 + \beta z^{-1}, 1)$ and $(z^{-1} + \beta z^{-2}, z^{-1})$ are both active.  Thus the dimension of the minimal transition space $\T_k$ of $\CC$ at any symbol time $k \in \Z$ is 2.

\pagebreak

More generally, let $\CC$ be any single-input, single-output LTI system whose shortest-generator set is the set of all shifts of the fundamental input-output trajectory $(b(z), a(z))$, whose delay is 0 and whose degree is $\delta = \max\{\deg b(z), \deg a(z)\}$;  then precisely $\delta + 1$ of these shifts are active at any symbol time $k \in \Z$, so the dimension  of the minimal transition space $\T_k$ of $\CC$ at any symbol time $k \in \Z$ is $\delta + 1$.
\qed


\vspace{1ex}
\noindent
\textbf{Example 2} (cont.).  Given the shortest basis $\{1111 0000,  0011 1100,  0000 1111, 0101 1010\}$ for the binary code $\CC$ of Example 2, we observe that the numbers of generators that are active at symbol times $0, 1, \ldots, 7$ are $1, 2, 3, 3, 3, 3, 2, 1$, respectively.  Thus these are the minimal transition space dimensions.  We say that the \emph{transition-space dimension profile} of $\CC$ is $\{1, 2, 3, 3, 3, 3, 2, 1\}$.
\qed \vspace{1ex}

In coding theory, the transition-space dimension profile of a linear code $\CC$ is generally taken as a better measure of the complexity of trellis-based decoding than its state-space dimension profile.

\subsection{Minimal realizations in controller canonical form}

Given any shortest basis $\B$ for any linear system $\CC$, we can now construct an obvious state-space realization for $\CC$, sometimes called the \emph{controller canonical form} \cite{Kailath}, which is evidently minimal. For multivariable LTI systems, a construction of a minimal realization in controller canonical form from a shortest (``minimal") basis was given in \cite{F75}.  In the literature of trellis realizations of block codes, such a construction was first given by Kschischang and Sorokine \cite{KS95}, who introduced the term ``atomic."

With each generator $\gb \in \B$, we associate an \emph{atomic} state-space realization as follows.    Roughly, it involves an ``input" $\alpha \in \F$ that occurs at symbol time $\del \gb$; a ``memory element" that stores $\alpha$ during the active state interval $(\del \gb, \deg \gb]$;  and an ``output" whose value during the active symbol interval $[\del \gb, \deg \gb]$ is $\alpha \gb$.

More precisely, if $\del \gb < \deg \gb$, then the state spaces of the atomic realization are equal to $\F$ during the active state interval $(\del \gb, \deg \gb]$, and equal to the trivial space $\{0\}$ otherwise;\footnote{
If $\del \gb = \deg \gb$, then $\Sigma_k = \{0\}$ for all $k \in \I_S$, and $\T_k = \{(0, \alpha g_k, 0): \alpha \in \F\}$ at symbol time $k = \del \gb = \deg \gb$;  otherwise $\T_k = \{(0, 0, 0)\}$.
}
thus the state space dimension is 1 during the active interval and 0 otherwise.  The sets of allowable transitions $\T_k$  are as given below during the active symbol interval $[\del \gb, \deg \gb]$ (otherwise $\T_k = \{(0, 0, 0)\}$):
\begin{itemize}  
\item For $k = \del \gb$, $\T_k = \{(0, \alpha g_k, \alpha): \alpha \in \F\}$;
\item For $\del \gb < k < \deg \gb$, $\T_k = \{(\alpha, \alpha g_k, \alpha): \alpha \in \F\}$;
\item For $k = \deg \gb$, $\T_k = \{(\alpha, \alpha g_k, 0): \alpha \in \F\}$;
\end{itemize}
thus the transition space dimension is 1 during the active interval, and 0 otherwise.   

The full behavior of this atomic realization is thus the one-dimensional vector space $\Bf = \{(\ab = \alpha \gb, \sigmab = \alpha \oneb_{(\del \gb, \deg \gb]}): \alpha \in \F\}$, where $\oneb_{(\del \gb, \deg \gb]}$ is the indicator function of the state interval $(\del \gb, \deg \gb]$.  The system that it realizes is the one-dimensional vector space $\{\alpha \gb: \alpha \in \F\}$;  \ie the subsystem of $\CC$ that is generated by $\gb$.

\pagebreak

The whole state-space realization for $\CC$ then consists of the aggregate of these atomic realizations for all $\gb \in \B$, plus an adder which at each symbol time produces the sum of the outputs of the currently active atomic realizations.  The set of all possible output trajectories of the whole realization is thus the set of all linear combinations $\sum_{\gb \in \B} \alpha(\gb) \gb$ of generators in $\B$, which is precisely the linear system $\CC$.  The number of memory elements active at any state time $k \in \I_S$ is the number of active generators at time $k$, which by the theorem above is the dimension of the minimal state space $\Sigma_k$ for $\CC$.  Thus this aggregate ``controller canonical form" realization is a minimal (and linear) realization of $\CC$.

In a controller canonical realization of a linear time-invariant system, the lengths of the generators $\gb \in \B$ are sometimes called the \emph{controllability indices} of $\CC$.  Thus in a linear time-varying system, the lengths of the generators may be regarded as generalized controllability indices. 

\subsection{New information, and forgetting information}

Further important quantities in a linear system $\CC$ are the amount of information that enters or ``drives" the system at each time, and also the amount that is ``forgotten" at each time.

We define the \emph{in-space} $I_k$ at symbol time $k \in \I$ as the quotient space $\CC_{k^+}/\CC_{(k+1)^+}$;  \ie the set of trajectories in $\CC$ that start at time $k$ or later, modulo those that start at time $k+1$ or later.

If $\B$ is a shortest basis for $\CC$, then $\CC_{k^+}$ is generated by the elements of $\B$ that have delay $k$ or more, and $\CC_{(k+1)^+}$ is generated by the elements of $\B$ that  have delay $k+1$ or more;  therefore:

\begin{theorem}[In-space dimension]
For any symbol time $k \in \I$, the dimension of the in-space $I_k$ of a linear system $\CC$ is the number of delay-$k$ generators in any shortest basis $\B$ for $\CC$.
\end{theorem}

As we have seen previously, the time-$k$ symbols of delay-$k$ generators in a shortest basis $\B$ must be linearly independent, so their number must satisfy $0 \le \dim I_k \le \dim A_k$.

Now if we compare the minimal state space $\Sigma_k = \CC/(\CC_{k^-} \times \CC_{k^+})$ to the minimal transition space $\T_k = \CC/(\CC_{k^-} \times \CC_{(k+1)^+})$, we see that 
$$
\dim \T_k = \dim \Sigma_k + \dim I_k.
$$
In words, if we regard one element of the field $\F$ as one unit of information, then in a minimal realization of $\CC$ the transition at symbol time $k$ is completely determined by $\dim \Sigma_k$ units of state information at state time $k$, plus $\dim I_k$ new units of information at symbol time $k$.  Indeed, in the controller canonical realization,  we see explicitly that the system transition at symbol time $k$ is completely determined by the coefficients of the $\dim \Sigma_k$ generators that are active at state time $k$, plus the $\dim I_k$ generators that start at symbol time $k$.

As an obvious corollary, if the minimal state spaces $\Sigma_k$ and the symbol alphabets $A_k$ are finite-dimensional, then the minimal transition spaces $\T_k$ are finite-dimensional.

Symmetrically, we may define the \emph{out-space} $O_k$ at symbol time $k \in \I$ as the quotient space $\CC_{(k+1)^-}/\CC_{k^-}$; then the dimension of $O_k$ is the number of degree-$k$ generators in any shortest basis $\B$ for $\CC$, and we have $\dim \T_k = \dim \Sigma_{k+1} + \dim O_k$, where $0 \le \dim O_k \le \dim A_k$.  In words, we may interpret $\dim O_k$ as the number of units of information that are ``forgotten" at symbol time $k$, as may be seen explicitly in the controller canonical realization.\footnote{Alternatively, if we were to run the system in the reverse-time direction, then $\dim O_k$ and $\dim I_k$ would reverse roles and become the amounts of ``new" and ``forgotten" information, respectively.}

\section{Minimal realizations of multivariable LTI systems}

We now consider minimal realizations of general Laurent LTI systems, which for brevity we  simply call \emph{multivariable LTI systems}.  The material in this section is well known;  our purpose is to let the reader see its connections with the ``shortest basis" approach.  

As usual, we make no distinction between $(\F((z^{-1})))^n$ and $\F^n((z^{-1}))$.


A multivariable LTI system is usually defined (as in Example 1) by an $n \times k$ rational \emph{generator matrix} $G(z) = \{g_{ij}(z): 1 \le i \le n, 1 \le j \le k\}$, where each element $g_{ij}(z)$ is a rational Laurent sequence.  A Laurent sequence in  $\F((z^{-1}))$ is \emph{rational} if it can be expressed in the form $a(z)/b(z)$, where $a(z)$ and $b(z)$ are polynomial sequences in $\F[z^{-1}]$ with $b(z) \neq 0$.  The set of all rational Laurent sequences is denoted by $\F(z^{-1})$, and forms a field;  in particular, the multiplicative inverse of a nonzero rational sequence $a(z)/b(z)$ is $b(z)/a(z)$.

The system $\CC$ is then the set of all $n$-tuples of Laurent sequences $\yb(z) = G(z) \ub(z)$ as $\ub(z)$ ranges through all $k$-tuples of Laurent sequences:
$$
\CC = \{\yb(z) =G(z) \ub(z) : \ub(z) \in (\F((z^{-1})))^k\}.
$$
(Notice that the component sequences $y_i(z)$ of $\yb(z)$ may represent either ``inputs" or ``outputs," as for example in Example 1.)  It follows that such an LTI system $\CC$ is a subspace of the $n$-dimensional vector space $(\F((z^{-1})))^n$ of all $n$-tuples of Laurent sequences over the Laurent field $\F((z^{-1}))$.  Thus without loss of generality we may assume that $k \le n$, and that $G(z)$ has full rank $k$.

We now give an ``algorithm" to find a shortest basis $\B$ for $\CC$.  The correctness of this approach is proved in $\cite{F70}$, using the invariant factor theorem (IFT), and again in $\cite{F75}$, without using the IFT.  The ``shortest basis" approach gives a nice motivation for this development.

We first find a set of $k$ independent finite generators whose shifts generate $\CC$, by finding the shortest finite trajectories in the $k$ 1-dimensional subsystems $\CC_j$ that are generated by the $k$ rational generators $\gb_j(z) = \{g_{ij}(z) = a_{ij}(z)/b_{ij}(z): 1 \le i \le n\}$.
We observe (as in Example 1) that $u_j(z)\gb_j(z)$ is finite if and only if $u_j(z)$ is a finite multiple of $\lambda_j(z)/\gamma_j(z)$, where $\lambda_j(z)$ is the least common multiple of the denominators $b_{ij}(z)$, and $\gamma_j(z)$ is the greatest common divisor of the numerators $a_{ij}(z)$.  The shortest finite trajectories in $\CC_j$ are therefore the shifts of $\gb'_j(z) = \lambda_j(z)\gb_j(z)/\gamma_j(z)$.    Since we may take any shift of $\gb'_j(z)$, we may assume without loss of generality that the delay of $\gb'_j(z)$ is zero.  Then the polynomial matrix  $G'(z) = \{\gb'_j(z): 1 \le j \le k\}$ is a basis for $\CC$.

We note in passing that this construction shows that any LTI system $\CC$ that is generated by a rational generator matrix $G(z)$ is controllable;  \ie generated by its finite trajectories.

Now we ask whether there exists any finite linear combination $G'(z)\ub(z)$ of the polynomial generators $\gb'_j(z)$ that produces a finite sequence that is shorter than any generator involved in this combination.  As we saw earlier, this can happen if and only if the set of shifts of the $\{\gb'_j(z)\}$ does not have the predictable delay property or the predictable degree property.  

\begin{enumerate}
\item The set of shifts of the $\{\gb'_j(z)\}$ does not have the predictable delay property if and only if there exists a linear combination $\gb''(z)$ of the delay-0 generators $\gb'_j(z)$ that has delay greater than zero.  This occurs if and only if the delay-0 coefficient $n$-tuples $\gb'_{j,0}$ are linearly dependent over $\F$, which occurs if and only if the $k \times k$ minors (determinants of $k \times k$ submatrices) of the $n \times k$ matrix $G'(z)$ all have delay greater than zero.  In this case we can obtain a shorter set of generators by substituting the linear combination $\gb''(z)$ for a longest generator $\gb'_j(z)$ that is involved in the combination.
\item The set of shifts of the $\{\gb'_j(z)\}$ does not have the predictable degree property if and only if there exists a linear combination $\gb''(z)$ of the degree-0 shifts $D^{-\deg \gb'_j(z)}\gb'_j(z)$ of the generators $\gb'_j(z)$ that has degree less than zero.  This occurs if and only if the high-order coefficient $n$-tuples $\gb'_{j, \deg \gb'_j}$ are linearly dependent over $\F$, which occurs if and only if the $k \times k$ minors of  $G'(z)$ all have degree less than their expected degree $\mu = \sum_{j=1}^k \deg \gb'_j(z)$.  In this case we can obtain a shorter set of generators by substituting the delay-0 shift of the linear combination $\gb''(z)$ for a longest generator $\gb'_j(z)$ that is involved in the combination.
\end{enumerate}

Finally, we ask whether the set of shifts of the $\{\gb'_j(z)\}$ is catastrophic--- \ie whether there exists any infinite linear combination $\yb(z) = G'(z)\ub(z)$ of the polynomial generators $\gb'_j(z)$ that produces a finite sequence $\yb(z)$.  As shown in \cite{F70, F75}, this occurs if and only if there is some polynomial $p(z) \in \F[z^{-1}]$ other than $z^{-1}$ and some finite $k$-tuple $\ub(z)$ such that $\gb''(z) = \yb(z)/p(z) = G'(z)(\ub(z)/p(z))$ is finite, whereas $\ub(z)/p(z)$ is infinite.  This occurs if and only if the matrix $G'(z) ~\mod p(z)$ has less than full rank, which occurs if and only if the $k \times k$ minors of  $G'(z)$ are all divisible by $p(z)$.   Then we can obtain a shorter set of generators by substituting the delay-0 shift of the linear combination $\gb''(z)$ for a longest generator $\gb'_j(z)$ that is involved in the combination.


To detect this situation, we can in principle compute the $k \times k$ minors of $G'(z)$, and see whether they have any common factor $p(z)$.  It turns out that $p(z)$ is an invariant factor of $G'(z)$ \cite{F70}, so any efficient algorithm for finding invariant factors of polynomial matrices may be used to find $p(z)$.  Then there exists some linear combination of the generators $\gb'_j(z)$ that equals zero modulo $p(z)$;  \ie is divisible by $p(z)$.  Dividing this combination by $p(z)$, we obtain our shorter generator $\gb''(z)$. 

\vspace{1ex}
\noindent
\textbf{Example 3}.  Consider the multivariable LTI system $\CC$ over any field $\F$ generated by the $3 \times 2$ matrix
$$
G(z) = \left[\begin{array}{cc}
1 & 1 - \alpha z^{-1} \\
1 - \beta z^{-1} & 1 \\
1 - \gamma z^{-1} & 1 - \delta z^{-1}
\end{array}\right],
$$
The $2 \times 2$ minors of $G(z)$ are $(\alpha + \beta) z^{-1} - \alpha \beta z^{-2}$,  $(\alpha + \gamma - \delta) z^{-1} - \alpha \gamma z^{-2}$, and $(\gamma - \beta - \delta) z^{-1} + \beta \delta z^{-2}$.  Since all have delay greater than zero (or, alternatively, since all have a common divisor $p(z) = z^{-1}$), the set of shifts of $\gb_1(z) = (1, 1 - \beta z^{-1}, 1 - \gamma z^{-1})$ and $\gb_2(z) = (1 - \alpha z^{-1}, 1, 1 - \delta z^{-1})$ does not have the predictable delay property.  Indeed, the linear combination $\gb_1(z) - \gb_2(z) = (\alpha z^{-1}, - \beta z^{-1}, (\delta - \gamma) z^{-1})$ has delay 1 and length 1, and its delay-0 shift $(\alpha, - \beta, (\delta - \gamma))$ (obtained by dividing out the common divisor $p(z) = z^{-1}$) may replace of $\gb_1(z)$ or $\gb_2(z)$ as a shorter fundamental generator.
\qed \vspace{1ex}

As this development suggests, the predictable delay, predictable degree, and non-catastrophic properties may be seen as special cases of the ``no common divisor" property, for the cases of $p(z) = z^{-1}$, $p(z) = z$, and all other polynomials, respectively; see the appendix of $\cite{F75}$.

When the ground field $\F$ is the complex field $\C$, then the minors of $G'(z)$ have no common divisor $p(z) \in \C[z^{-1}]$ if and only if they have no common degree-1 divisor $z^{-1} - \alpha$ for any $\alpha \in \C$.  As Example 3 illustrates, the case $\alpha = 0$ corresponds to a test of the predictable delay property;  by interchanging $z$ and $z^{-1}$ we can see that the case $\alpha = \infty$ ($\alpha^{-1} = 0$) corresponds to a test of the predictable degree property. 
Since $G'(z) ~\mod z^{-1} - \alpha$ is the complex matrix obtained by ``evaluating" $G'(z)$ at $z^{-1} = \alpha$, this ultimately implies that $G'(z)$ is a set of shortest fundamental generators for $\CC$ if and only if $G'(z)$ has full rank when evaluated at $z^{-1} = \alpha$ for all $\alpha \in \C \cup \infty$.  In system theory, this property is sometimes expressed in terms like the following:  ``The matrix $G'(z)$ has no zeroes anywhere in the complex plane, including at zero and at infinity."  \pagebreak

\section{Duality}

An alternative way of defining a linear system $\CC$ is via a set of generators for its orthogonal system $\CC^\perp$.  In linear system theory, such a representation of $\CC$ is sometimes called a \emph{kernel representation}, whereas a representation in terms of generators for $\CC$ is called an \emph{image representation}.

If $\B^\perp$ is a shortest basis for $\CC^\perp$, then $\CC$ is the set of all trajectories that are orthogonal to all trajectories in $\B^\perp$.  A fundamental duality result is that any minimal state space for $\CC^\perp$ has the same dimension as the corresponding minimal state space for $\CC$.  These results lead to a minimal realization for $\CC$ in ``observer canonical form" \cite{Kailath}.  We also determine the dimensions of the minimal transition spaces of $\CC^\perp$.

\subsection{Orthogonal systems}

Each symbol alphabet $A_k$ is a finite-dimensional vector space over $\F$, and therefore has a dual space $\hat{A}_k$ of the same dimension such that for all $a_k \in A_k, \hat{a}_k \in \hat{A}_k$ there is a well-defined inner product $\inner{a_k}{\hat{a}_k} \in \F$.  Commonly $A_k$ is the set $\F^n$ of $n$-tuples over $\F$;  then $\hat{A}_k$ may also be taken as $\F^n$, with the inner product being defined in standard componentwise fashion.


If $\A = \prod_{k \in \I} A_k$ is the set of all Laurent trajectories, then its dual space $\hat{\A} = \prod_{k \in \I} \hat{A}_k$ is the set of all anti-Laurent trajectories (\ie all nonzero trajectories with finite degree, plus $\zerob$). 
The inner product between a trajectory $\ab \in  \A$ and a dual trajectory $\hat{\ab} \in  \hat{\A}$ may then be defined by $\inner{\ab}{\hat{\ab}} = \sum_{k\in\I} \inner{a_k}{\hat{a}_{k}}$;  this sum is well defined if $\ab$ is Laurent and $\hat{\ab}$ is anti-Laurent, because then only a finite number of terms in the sum are nonzero.

The \emph{orthogonal system} $\CC^\perp$ is then defined as set of all trajectories $\hat{\ab} \in \hat{\A}$ whose inner product  with all trajectories $\ab \in \CC$ is zero.  $\CC^\perp$ is linear, and its orthogonal system $\CC^{\perp\perp}$ is $\CC$.  Therefore $\CC$ may be characterized as the set of all $\ab \in \A$ that are orthogonal to a basis $\B^\perp$ for $\CC^\perp$.

\vspace{1ex}
\noindent
\textbf{Example 2} (cont.).  The orthogonal code $\CC^\perp$ to the code $\CC$ of Example 2 is $\CC$ itself.  Thus an 8-tuple $\ab \in (\F_2)^8$ is in $\CC$ if and only if it is orthogonal to the four generators $\gb_1, \gb_2, \gb_3, \gb_4$. \qed

\vspace{1ex}

If a linear system $\CC$ is Laurent complete (the Laurent completion of its finite subcode), then $\CC^\perp$ is anti-Laurent complete, so there is a nice symmetry between a system and its orthogonal system.\footnote{This symmetry can be improved by reversing the time axis of the orthogonal code.}

An LTI system $\CC$ over a field $\F$ is invariant both under multiplication by elements of $\F$ and under time shifts, and therefore is invariant under multiplication by Laurent sequences in $\F((D))$ or $\F((z^{-1}))$, where multiplication of Laurent trajectories by Laurent sequences is defined by sequence convolution. Thus we usually prefer to regard an LTI system as a linear system over $\F((D))$ or $\F((z^{-1}))$.  In this case the orthogonal code $\CC^\perp$ is defined by an inner product in $\F((D))$ or $\F((z^{-1}))$, and is also linear over $\F((D))$ or $\F((z^{-1}))$.  It is easy to see that the orthogonal code $\CC^\perp$ so defined is the time reversal of the orthogonal code according to the earlier definition using an inner product over $\F$.  Thus we use slightly different definitions of orthogonality for LTI and for non-LTI systems.\footnote{As noted in the previous footnote, it would be better conceptually if we always reversed the time axis of the orthogonal system $\CC^\perp$, but in the general case the notation then becomes cumbersome.}

\vspace{1ex}
\noindent
\textbf{Example 1} (cont.).  The linear time-invariant system $\CC$ of Example 1 may be regarded as the one-dimensional subspace of the two-dimensional vector space $(\F((z^{-1})))^2$ that is generated by $(1, g(z))$.  The orthogonal code $\CC^\perp$ is then the orthogonal one-dimensional subspace, which is generated by $(g(z), -1)$.  In other words, a pair $(u(z), y(z))$ is in $\CC$ if and only if $u(z)g(z) - y(z) = 0$. \qed

\subsection{Dual minimal state and transition spaces}

It is well known that the minimal state spaces $\hat{\Sigma}_{k}$ of the orthogonal system $\CC^\perp$ have the same dimensions as the corresponding minimal state spaces $\Sigma_k$ of $\CC$.  This follows from:
\begin{itemize}
\item The fact that $\Sigma_k$ is isomorphic to the past-induced state space $P_{k^-}(\CC)/\CC_{k^-}$, or equivalently to the quotient space $R_{k^-}(\CC)/R_{k^-}(\CC_{k^-})$, where the restriction $R_{k^-}$ maps a trajectory $\ab = \{a_{k'}: k' \in \I\}$ defined on $\I$ to a restricted trajectory $R_{k^-}(\ab) = \{a_{k'},:k' \in k^-\}$ defined on $k^-$.
\item The fact that restricted systems and subsystems are duals, in the sense that $R_{k^-}(\CC^\perp) = (R_{k^-}(\CC_{k^-}))^\perp$ and $R_{k^-}((\CC^\perp)_{k^-}) = (R_{k^-}(\CC))^\perp$.  This follows from the fact that $(R_{k^-}(\ab), R_{k^+}(\ab))$ is orthogonal to $(R_{k^-}(\bb), R_{k^+}(\zerob))$ if and only if $R_{k^-}(\ab)$ is orthogonal to $R_{k^-}(\bb)$.
\item The fact that if $B$ is a subspace of $A$, then $A^\perp$ is a subspace of $B^\perp$, and the quotient spaces $A/B$ and $B^\perp/A^\perp$ are dual spaces, with the same dimension.  Consequently $R_{k^-}(\CC)/R_{k^-}(\CC_{k^-}) \simeq \Sigma_k$ and $R_{k^-}(\CC^\perp)/R_{k^-}((\CC^\perp)_{k^-}) \simeq \hat{\Sigma}_k$ are dual spaces, and
$\dim \Sigma_k = \dim \hat{\Sigma}_k$.
\end{itemize}

Is there a corresponding duality result for minimal transition spaces?  A little-known fact, apparently first discovered by Mittelholzer \cite{M95}, and proved in a more general context in \cite{F01}, is that whereas $\Sigma_k$ and $\hat{\Sigma}_k$ are dual spaces, if $\T_k \subseteq \Sigma_k \times A_k \times \Sigma_{k+1}$ is a minimal transition space of $\CC$, then the corresponding minimal transition space of $\CC^\perp$ is the orthogonal space $(\T_k)^\perp$ to $\T_k$ in the dual space $\hat{\Sigma}_{k} \times \hat{A}_{k} \times \hat{\Sigma}_{k+1}$, where orthogonality is defined with respect to the following bilinear form:
$$
\inner{(\sigma_k, a_k, \sigma_{k+1})}{(\hat{\sigma}_k, \hat{a}_k, \hat{\sigma}_{k+1})} = \inner{\sigma_k}{\hat{\sigma}_k} + \inner{a_k}{\hat{a}_k} - \inner{\sigma_{k+1}}{\hat{\sigma}_{k+1}}.
$$
Thus the dimensions of the minimal transition spaces of $\CC$ and $\CC^\perp$ are in general related as follows:
$$
\dim \T_k + \dim (\T_{k})^\perp = \dim \Sigma_k + \dim A_k + \dim \Sigma_{k+1}.
$$

The in-space of $\CC^\perp$ at time $k$ is in some sense the dual to the out-space of $\CC$ at time $k$, and \emph{vice versa}, as can be seen from the following relations.
Since $\dim \Sigma_k = \dim \hat{\Sigma}_k$ and $\dim \T_k = \dim \Sigma_k + \dim I_k = \dim \Sigma_{k+1} + \dim O_k$, we have
$$
\dim (\T_{k})^\perp = \dim \hat{\Sigma}_k + \dim A_k - \dim O_k = \dim \hat{\Sigma}_{k+1} + \dim A_k - \dim I_k.
$$
Thus if $\hat{I}_k$ and $\hat{O}_k$ are the in-space and out-space of $\CC^\perp$ at symbol time $k$, then
\begin{eqnarray*}
\dim \hat{I}_k & = & \dim A_k - \dim O_k; \\
\dim \hat{O}_k & = & \dim A_k - \dim I_k.
\end{eqnarray*}
In particular, when symbols are simply elements of $\F$ so that  $\dim A_k = 1$, then $\dim \hat{O}_k = 0$ if $\dim I_k = 1$ and \emph{vice versa}.

\vspace{1ex}
\noindent
\textbf{Example 1} (cont.). Let $\CC$ be any single-input, single-output linear time-invariant system, whose shortest basis is the set of all shifts of the fundamental input-output trajectory $(b(z), a(z))$, whose delay is 0 and whose degree is $\delta = \max\{\deg b(z), \deg a(z)\}$.  Then a shortest basis for the orthogonal system $\CC^\perp$ is the set of all shifts of the fundamental input-output trajectory $(a(z), -b(z))$.  Thus the minimal state spaces of $\CC$ and $\CC^\perp$ both have dimension $\delta$, and the minimal transition spaces of $\CC$ and $\CC^\perp$ both have dimension $\delta +1$.  Also, all in-spaces and out-spaces have dimension 1.  Since $\dim A_k = 2$, it is easy to check all that the relations above are satisfied.  \qed

\vspace{1ex}
\noindent
\textbf{Example 2} (cont.).  The binary code $\CC$ of Example 2 is generated by the shortest basis $\{1111 0000$, $0011 1100,  0000 1111, 0101 1010\}$, and has state-space dimension profile $\{0, 1, 2, 3, 2, 3, 2, 1, 0\}$ and transition-space dimension profile $\{1, 2, 3, 3, 3, 3, 2, 1\}$.  The orthogonal code $\CC^\perp$ is the same code, and thus has the same profiles.  One may check that the relations above are satisfied at all times;  for example, at symbol time 0, $\dim I_0 =  \dim \hat{I}_0 = 1, \dim O_0 = \dim \hat{O}_0 = 0$ and $\dim \T_0  = \dim  (\T_{0})^\perp = 1$.  Note that since $\dim A_k = 1$ and $\CC = \CC^\perp$, we must have $\dim O_k = 0$ if $\dim I_k = 1$, and \emph{vice versa}; thus $\dim I_k = 1$ for $k = 0, 1, 2, 4$, while $\dim O_k = 1$ for $k = 3, 5, 6, 7$.  
\qed \vspace{1ex}

\noindent
\textbf{Example 3} (cont.).  The system $\CC$ of Example 3 is a two-dimensional subspace of $(\F((z^{-1})))^3$, and has a shortest basis consisting of the shifts of the two delay-0 generators $\gb'_1(z) = (\alpha, - \beta, (\delta - \gamma))$ and $\gb_2(z) = (1 - \alpha z^{-1}, 1, 1 - \delta z^{-1})$, which have degrees 0 and 1, respectively.  Thus the minimal state spaces of $\CC$ have dimension 1, and its minimal transition spaces have dimension 3.  The orthogonal system $\CC^\perp$ must thus be a one-dimensional subspace of $(\F((z^{-1})))^3$, and must be generated by the shifts of a delay-0, degree-1 generator $\hb(z)$. Thus its minimal state spaces have dimension 1, and its minimal transition spaces have dimension 2.
\qed \vspace{1ex}

More generally, as we have seen, a multivariable LTI system $\CC$ has a polynomial $n \times k$ generator matrix $G'(z) = \{\gb'_j(z): 1 \le j \le k\}$, whose $k \times k$ minors have no common polynomial factors, and have maximum degree $\mu = \sum_{j = 1}^k \deg \gb'_j(z)$;  then the shifts of the fundamental generators $\gb'_j(z)$ form a shortest basis for $\CC$.  It turns out that the orthogonal LTI system $\CC^\perp$ has a polynomial $n \times (n-k)$ generator matrix $H(z) = \{\hb_j(z): 1 \le j \le n - k\}$ whose $(n - k) \times (n-k)$ minors are the same (up to sign) as the complementary $k \times k$ minors of $G'(z)$ \cite{F73}, such that the shifts of the fundamental generators $\hb_j(z)$ form a shortest basis for $\CC^\perp$.  Since the in-spaces and out-spaces of $\CC$ have dimension $k$, and those of $\CC^\perp$ have dimension $n-k$, it follows that the dimensions of the minimal state spaces of $\CC$ and $\CC^\perp$ are both equal to $\mu$, that the dimensions of the minimal transition spaces of $\CC$ are $\mu + k$, and that the dimensions of the minimal transition spaces of $\CC^\perp$ are $\mu + n-k$.

\vspace{1ex}
\noindent
\textbf{Example 3} (cont.).  For the system $\CC$ of Example 3, the $2 \times 2$ minors of $G'(z)$ are $\alpha + \beta - \alpha \beta z^{-1}$,  $\alpha + \gamma - \delta - \alpha \gamma z^{-1}$, and $\gamma - \beta - \delta + \beta \delta z^{-1}$.  It follows that these are also the $1 \times 1$ minors of the  generator matrix $H(z)$ of $\CC^\perp$, up to sign;  indeed, the fundamental generator of $H(z)$ is $\hb(z) = (\gamma - \beta - \delta + \beta \delta z^{-1}, -\alpha - \gamma + \delta + \alpha \gamma z^{-1}, \alpha + \beta - \alpha \beta z^{-1})$.
\qed 

\subsection{Minimal realizations in observer canonical form}

Given a shortest basis $\B^\perp$ for the orthogonal system $\CC^\perp$ to a linear system $\CC$, we can give a straightforward state-space realization for $\CC$, sometimes called the \emph{observer canonical form} \cite{Kailath}, whose state-space dimensions are minimal at all times.  Indeed, this realization may be obtained by dualizing the controller canonical form realization of $\CC^\perp$.

With each generator $\hb \in \B^\perp$, we associate a one-dimensional atomic ``checker" as follows.    Roughly, for an arbitrary trajectory $\ab \in \A$, the realization accumulates the partial sums of the inner product $\inner{\ab}{\hb} = \sum_{k \in [\del \hb, \deg \hb]} \inner{a_{k}}{h_k}$ in an ``accumulator."  The trajectory ``checks" (is valid) with respect to $\hb$ if the final sum in the accumulator is 0.

More precisely, if $\del \hb < \deg \hb$, then the state spaces of the atomic checker are equal to $\F$ during the active state interval $(\del \hb, \deg \hb]$, and equal to the trivial space $\{0\}$ otherwise;\footnote{
If $\del \hb = \deg \hb$, then $\hat{\Sigma}_k = \{0\}$ for all $k \in \I_S$, and $\T_k = \{(0,  a_k, 0) : \inner{a_k}{h_k} = 0\}$ at symbol time $k = \del \hb = \deg \hb$;  otherwise $\T_k = \{(0, 0, 0)\}$.
}
 thus the state space dimension is 1 during the active interval and 0 otherwise.  The sets of allowable transitions $\T_k$  are as given below during the active symbol interval $[\del \hb, \deg \hb]$ (otherwise $\T_k = \{(0, 0, 0)\}$):

\begin{itemize}  
\item For $k = \del \hb$, $\T_k = \{(0, a_{k}, \sigma_{k+1} = \inner{a_{k}}{h_{k}}): a_k \in A_k\}$;
\item For $\del \hb < k < \deg \hb$, $\T_k = \{(\sigma_k, a_{k}, \sigma_{k+1} = \sigma_k + \inner{a_{k}}{h_{k}}): a_k \in A_k\}$;
\item For $k = \deg \hb$, $\T_k = \{(-\inner{a_{k}}{h_k}, a_{k}, 0): a_k \in A_k\}$.
\end{itemize}  
(It is easily checked that these are the orthogonal transition spaces under the bilinear form given above to those defined in Section IV-D for the controller canonical form, if  $\hb$ is substituted for $\gb$.)
Evidently $\sigma_k = \sum_{k' < k} \inner{a_{k'}}{h_{k'}}$ for $k \le \deg \hb$, and  a trajectory $\ab \in \A$ has a corresponding state sequence $\sigmab(\ab)$ such that $(\ab, \sigmab(\ab))$ is a valid symbol-state trajectory if and only if 
$$
\sigma_{\deg \hb} = \sum_{k \in [\del \hb, \deg \hb)} \inner{a_{k}}{h_k} = -\inner{a_{\deg \hb}}{h_{\deg \hb}};
$$
 \ie if and only if $\inner{\ab}{\hb} = 0$.

The whole realization for $\CC$ then consists of the aggregate of these atomic ``checkers" for all $\hb \in \B^\perp$.  The set of all $\ab \in \A$ that have a compatible state sequence $\sigmab(\ab)$ in all ``checkers" is the set of all $\ab \in \A$ that are orthogonal to all $\hb \in \B^\perp$, which is precisely the linear system $\CC$.  The number of memory elements active at any state time $k \in \I_S$ is the number of active $\hb \in \B^\perp$ at time $k$, which is the dimension of the minimal state space $\hat{\Sigma}_{k}$ for $\CC^\perp$, which equals $\dim \Sigma_k$.  Thus this aggregate ``observer canonical form" realization is a minimal (and linear) realization of $\CC$.

In an observer canonical realization of a linear time-invariant system, the lengths of the dual generators $\hb \in \B^\perp$ are sometimes called the \emph{observability indices} of $\CC$.  Thus the observability indices of $\CC$ are the controllability indices of $\CC^\perp$, and \emph{vice versa}. 

\section{Conclusion}

The ``shortest basis" approach describes a linear system $\CC$ by a shortest basis $\B$ that transparently characterizes the controllability properties of $\CC$, or by a shortest basis $\B^\perp$ for $\CC^\perp$ that similarly characterizes the observability properties of $\CC$.  In particular, $\B$ and $\B^\perp$ lead directly to minimal linear realizations of $\CC$ in controller and observer canonical form, respectively.  

In this paper, we have not carried this approach through to the development of actual algorithms for computing minimal realizations, given a description of a linear system $\CC$.  When the time axis $\I$ is finite, then there exist straightforward algorithms for reducing a given basis $\B$ to a shortest basis by detecting and correcting any failure of $\B$ to have the predictable span property;  see \cite{V98}.  When the time axis is infinite, we need also to detect and correct catastrophicity.  For LTI systems, this is essentially a matter of finding and eliminating invariant polynomial factors, as discussed in Section V.  For linear time-varying systems, detecting catastrophicity is in general an open question, depending on how the system is described.

As shown in \cite{FT93}, this approach generalizes naturally to discrete-time group systems.  As shown in \cite{F01}, it further generalizes to linear and group systems defined on cycle-free graphs, rather than on a standard discrete time axis, and to some extent to general graphs.  Such generality suggests that the ``shortest basis" approach is rather fundamental. 

\section*{Acknowledgments}  I have benefitted greatly from interactions with many colleagues over the very many years during which I have been interested in these topics, including particularly F. Fagnani,  R. Johannesson, T. Kailath, R. Koetter, F. R. Kschischang, H.-A. Loeliger, J. L. Massey, T. Mittelholzer, S. Mitter, J. Rosenthal, M. D. Trott,  A. Vardy, J. C. Willems, and S. Zampieri.

For comments on earlier versions of this paper, I am grateful to H. Gluesing-Luerssen, F. R. Kschischang, M. Kuijper, and the reviewers.

\section*{Appendix.  The shortest basis approach for complete systems}

In this Appendix, we relax the restriction that trajectories $\ab \in \A = \prod_{k\in\I} A_k$ must be Laurent.  This allows us to consider uncontrollable systems $\CC \subset \A$, which in general will have autonomous components.  The theory is a straightforward extension of the approach developed for controllable systems in the main body of the paper, but now a shortest basis $\B$ may include infinite generators.  However, we will see that the duality theory becomes less symmetrical, since the dual space $\hat{\A}$ now consists of only the finite trajectories in $\prod_{k\in\I} \hat{A}_k$.

We correspondingly relax the restriction that linear combinations must be Laurent.  Thus, if $\B$ is a basis for $\CC$, then $\CC$ is the set of \emph{all} linear combinations of generators in $\B$.  This implies that $\CC$ is ``complete," not just ``Laurent complete." 

We continue to define a system $\CC$ as \emph{controllable} if it is generated by its finite trajectories, where ``generated" is now understood in the sense of unrestricted linear combinations.  A complete LTI system need not be controllable, as is shown by the following simple example.

 \vspace{1ex}
\noindent
\textbf{Example A} (repetition system).  Over any field $\F$, the repetition system $\CC$ is defined as the set of all bi-infinite trajectories $\ab \in \F^\Z$ such that $a_k = \alpha$ for all $k \in \Z$, for some $\alpha \in \F$.  $\CC$ is evidently a linear time-invariant system, and indeed a one-dimensional subspace of $\F^\Z$ that is generated by the bi-infinite all-one trajectory $\oneb \in \F^\Z$.  Since the only finite trajectory in $\CC$ is the all-zero trajectory $\zerob$, $\CC$ is uncontrollable.
\qed \vspace{1ex}

The controllable subsystem of a linear system $\CC$ is defined as the subsystem $\CC^c$ generated by all finite trajectories of $\CC$.  The uncontrollable component of $\CC$ may be defined abstractly as the quotient space $\CC/\CC^c$.

As is well known, the minimal state space theorem continues to hold for uncontrollable systems, and it is easy to see that the minimal transition space theorem does also.    Furthermore, Theorems 4 and 5 continue to hold;  \ie the minimal state space (resp.\ transition space) dimension at state (resp.\ symbol) time $k$ is the number of active generators at time $k$. Thus if we continue to require that minimal state spaces be finite-dimensional, then the dimension of $\CC^u$ must be finite.
Thus $\CC^u$ must have a basis $\B^u$ consisting of a finite set of infinite trajectories in $\CC$ (\eg the all-one trajectory $\oneb$ in Example A).  


For the controllable subsystem $\CC^c$, the ``shortest basis" approach of the main body of the text goes through without significant change.  In particular, we may still obtain a basis $\B^c$ for $\CC^c$ by our greedy construction.  This construction may be continued to construct a shortest basis $\B^u$ for $\CC^u$ consisting of a finite set of infinite generators.  Together, $\B^c$ and $\B^u$ form a shortest basis $\B$ for $\CC$.

Minimal linear realizations in controller canonical form may be constructed as before, with the only difference being that some of the atomic generators may be active for an infinitely long time.

 \vspace{1ex}
\noindent
\textbf{Example A} (cont.).  For a repetition system $\CC$, the subsystems $\CC_{k^-}$ and $\CC_{k^+}$ are trivial at all times $k \in \Z$, so the minimal state space and transition space have dimension 1 at all times.  Any nonzero trajectory in $\CC$ may be taken as the single generator in a shortest basis $\B$;  such a generator is active at all times.  A minimal realization of $\CC$ in controller canonical form is given by the one-dimensional atomic realization whose transition space is $\T_k = \{(\alpha, \alpha, \alpha): \alpha \in \F\}$ at all times.
\qed \vspace{1ex}

It turns out that the dual space to the complete space $\prod_{k\in\I} A_k$ is the space  $\hat{\A}$ consisting of the finite trajectories in $\prod_{k\in\I} \hat{A}_k$.  This ensures that the inner product $\inner{\ab}{\hat{\ab}}$ of a trajectory $\ab \in \A$ and a trajectory $\hat{\ab} \in \hat{\A}$ is well defined, since such an inner product is a sum of only finitely many nonzero terms $\inner{a_k}{\hat{a}_k}$.  Linear combinations in $\hat{A}$ are restricted to finite linear combinations.

Consequently, the orthogonal system $\CC^\perp \subseteq \hat{\A}$ to a complete system $\CC \subseteq \A$ is a finite system;  \ie a system all of whose trajectories are finite.  Such a system is necessarily controllable--- \ie generated as the set of all finite linear combinations of its (finite) elements.  
Therefore the shortest basis approach of the main body of this paper applies directly to $\CC^\perp$.

 \vspace{1ex}
\noindent
\textbf{Example A} (cont.).  The orthogonal system $\CC^\perp$ to a repetition system $\CC$ is the set of all finite trajectories $\hat{\ab} \in \F^\Z$ that are orthogonal to the all-one trajectory $\oneb$;  \ie the set of all finite trajectories $\hat{\ab} \in \F^\Z$ whose components sum to zero:  $\sum_{k\in\I} \hat{a}_k = 0$.  
Note that $\CC^\perp$ is also a time-invariant system.

Since a zero-sum trajectory cannot have length 1, the shortest nonzero trajectories in $\CC^\perp$ have length 2 and are of the form $\alpha(D^k - D^{k+1})$ for some $\alpha \in \F$ and $k \in \Z$.  The set of all shifts of the fundamental generator $1-D$ generates $\CC^\perp$, and is therefore a shortest basis $\B$ for $\CC^\perp$.
\qed \vspace{1ex}

Our duality results for minimal state spaces and transition spaces of $\CC^\perp$ continue to hold.  From a shortest basis for $\CC^\perp$, we can construct a minimal realization of $\CC^\perp$ in controller canonical form, or a minimal realization of $\CC$ in observer canonical form.

 \vspace{1ex}
\noindent
\textbf{Example A} (cont.).  The elements $D^k - D^{k+1}$ of a shortest basis $\B$ for $\CC^\perp$ all have length 2.  One is active at each state time, and two are active at each symbol time.  Therefore the minimal state space dimension is 1 at each time, the same as the minimal state space dimension for $\CC$, and the minimal transition space dimension is 2, which equals 
$$
\dim \Sigma_k + \dim A_k + \dim \Sigma_{k+1} - \dim \T_k = 1 + 1 + 1 - 1.
$$
The in-space and out-space dimensions of $\CC$ are 0 at all times, and those of $\CC^\perp$ are 1 at all times.  

A minimal realization for $\CC^\perp$ in controller canonical form may be constructed from an infinite set  of one-dimensional atomic realizations, one for each $k \in \Z$, of which the $k$th produces an output sequence $\alpha(D^k - D^{k+1})$ and is active only at one state time, namely $k+1$.  Similarly, a minimal realization for $\CC$ in observer canonical form may be constructed from an infinite set  of one-dimensional atomic checkers, one for each $k \in \Z$, of which the $k$th checks that the inner product of $\ab$ with $D^k - D^{k+1}$ is zero, and is active only at state time $k+1$.
\qed \vspace{1ex}

In summary, the shortest basis approach applies equally well to complete systems.

\pagebreak

\pagebreak

\section*{Author biography}

	\textbf{G. David Forney, Jr.} received the B.S.E. degree in electrical engineering from Princeton University, Princeton, NJ, in 1961, and the M.S. and Sc.D. degrees in electrical engineering from the Massachusetts Institute of Technology, Cambridge, MA, in 1963 and 1965, respectively.

	From 1965-99 he was with the Codex Corporation, which was acquired by Motorola, Inc. in 1977, and its successor, the Motorola Information Systems Group, Mansfield, MA.  Since 1996, he has been an Adjunct Professor at M.I.T.

	Dr. Forney was Editor of the IEEE Transactions on Information Theory from 1970 to 1973.  He has been a member of the Board of Governors of the IEEE Information Theory Society during 1970-76, 1986-94, and 2004-10, and was President in 1992 and 2008.  He has been awarded the 1970 IEEE Information Theory Group Prize Paper Award, the 1972 IEEE Browder J. Thompson Memorial Prize Paper Award, the 1990 and 2009 IEEE Donald G. Fink Prize Paper Awards, the 1992 IEEE Edison Medal, the 1995 IEEE Information Theory Society Claude E. Shannon Award, the 1996 Christopher Columbus International Communications Award, and the 1997 Marconi International Fellowship.  In 1998 he received an IT Golden Jubilee Award for Technological Innovation, and two IT Golden Jubilee Paper Awards.  He received an honorary doctorate from EPFL, Lausanne, Switzerland in 2007.  He was elected a Fellow of the IEEE in 1973, a member of the National Academy of Engineering (U.S.A.) in 1983, a Fellow of the American Association for the Advancement of Science in 1993, an honorary member of the Popov Society (Russia) in 1994, a Fellow of the American Academy of Arts and Sciences in 1998, and a member of the National Academy of Sciences (U.S.A.) in 2003.

\end{document}